\documentstyle[prd,aps,epsfig,preprint,tighten]{revtex}
\begin{document}
\preprint{                                                 BARI-TH/273-97}
\draft
\title{ Reconciling solar and terrestrial neutrino oscillation\\ 
			evidences with minimum sacrifice}
\author{         G.~L.~Fogli, E.~Lisi, D.~Montanino, and G.~Scioscia}
\address{     Dipartimento di Fisica and Sezione INFN di Bari,\\
                  Via Amendola 173, I-70126 Bari, Italy}
\maketitle
\begin{abstract}
The present possible evidences in favor of neutrino masses and mixings from 
solar, atmospheric, and accelerator experiments cannot be all reconciled 
in a three-family framework, unless some data are excluded. We grade
all possible three-family scenarios according to their compatibility
with the available data. A recently proposed scenario appears to emerge 
naturally as the most likely solution to all  oscillation evidences, with 
the  only exception of the  angular dependence of multi-GeV atmospheric 
data in the  Kamiokande experiment. We describe in detail the status and the 
phenomenological implications of this ``minimum sacrifice'' solution.
\end{abstract}

\section{Introduction}

	At present, there are three possible evidences for neutrino
masses and mixings: the solar neutrino problem  \cite{Sola}, the atmospheric
neutrino anomaly \cite{Atmo}, and the event excess in the Liquid Scintillator 
Neutrino Detector (LSND) experiment  \cite{LSND}. These evidences
are individually best-fitted by three largely different mass gaps in
the neutrino spectrum, and thus cannot be all reconciled in a
three-flavor ($3\nu$) oscillation framework, unless some data are excluded
(see, e.g., \cite{Sm96}).

	In this work we grade all possible three-flavor scenarios according
to their compatibility with the world neutrino data. The specific data  
conflicting with the various scenarios are systematically identified.
We treat in increasing detail those $3\nu$ frameworks that require 
a decreasing ``sacrifice'' of data in order to achieve a good global fit. 
The scheme recently proposed by Cardall and Fuller \cite{Ca97} will emerge
as the ``minimum sacrifice'' scenario, and its implications for current 
and future experiments will be investigated in detail. In developing our 
analysis, we also  comment on other  three-flavor analyses recently 
appeared in the literature \cite{Mi95,Go95,Ac96,Ha97}.
This work builds upon  our previous studies of  solar \cite{Fo96,Fa97},
atmospheric \cite{Li95,Fo97}, and laboratory (accelerator and
reactor) \cite{Fo95,Li97} neutrino oscillations, to which the reader is 
referred for further details and extensive bibliography.

	The paper is organized as follows. In Sec.~II we discuss briefly the 
evidences for neutrino masses and mixings and introduce the notation for 
three-flavor oscillations. In Sec.~III we discuss both  hierarchical and 
non-hierarchical scenarios, and conclude that the latter are globally
disfavored. In Sec.~IV we  identify a specific hierarchical framework that 
demands the ``minimum sacrifice'' of data.  Its implications for current and 
future experiments are investigated in Sec.~V. We  draw the conclusions of 
our work in Sec.~VI.

\section{Neutrino masses and mixings: experimental evidences and
three-flavor notation}

\subsection{Evidences for neutrino mixing and oscillation}

	It is useful to start by making a  distinction between evidences for
neutrino {\em mixing\/} and for neutrino {\em oscillation\/}. We define 
``evidences for $\nu$ mixing''  those signals  of flavor transitions, either 
direct (flavor appearance) or indirect (flavor disappearance),
that imply  nonzero mixing but do not provide a measurement of the oscillation 
wavelength. We then promote these basic signals  to ``evidences for $\nu$ 
oscillation''  only if, in addition, an effect related the periodicity
of the flavor transition process (i.e., to  the oscillation wavelength) is 
detected. In terms of neutrino events, evidences for {\em mixing\/} are 
simply established by ``anomalous'' measurements of {\em total\/}
neutrino rates, while evidences for  {\em oscillation\/} require, in addition,
the more delicate observation of anomalous {\em spectra\/} of events as a
function of either the neutrino path length $L$ or the neutrino energy $E$.

	The implications on the neutrino mass square difference $\Delta m^2$ 
(assuming for the moment two neutrino families) are rather different
in the two cases. An evidence for $\nu$ mixing can only set a trivial lower 
bound on $\Delta m^2$, below which the oscillation length would largely 
overshoot the experimental baseline $L$ and flavor transitions would not 
have time to develop. An evidence for neutrino oscillations, instead, can 
give more detailed indications on $\Delta m^2$ through the measured $E$ or $L$
spectra, since $\Delta m^2$ appears either in the combination 
$\Delta m^2\times L/E$ (vacuum oscillations) or $\Delta m^2/E$ 
(matter oscillations).

	At present, there are three primary evidences for neutrino 
{\em mixing}: the overall deficit of the solar neutrino flux \cite{Sola}, 
the anomalous flavor ratio of total atmospheric rates \cite{Atmo}, and the 
LSND event excess \cite{LSND}.  For the sake of brevity, they will be termed 
S (for solar), A (for atmospheric), and L (for LSND, laboratory), respectively. 
The following additional (and comparatively less established)  observations 
may suggest genuine  neutrino {\em oscillation\/} effects:
(${\rm S}'$) the solar neutrino deficit in different experiments \cite{Sexp}
appears to depend significantly on the neutrino energy $E$ (see \cite{KrPe} 
and references therein); (${\rm A}'$) the atmospheric 
anomaly of multi-GeV neutrinos in Kamiokande seems to depend on the zenith
angle (i.e., on  the path length $L$) \cite{Fu94}, although with uncertain 
statistical significance \cite{Li95,Sa95}; and (${\rm L}'$) there is some 
evidence for a peak in the $L/E$ spectrum of LSND events 
(see Fig.~32 in \cite{LSND}).

	Tables I and II list the above evidences for neutrino {\em mixing\/}
and {\em oscillation}, respectively, together with their implications
in terms of $\Delta m^2$. Notice that the evidences S, A, and L,
are all compatible with large values of $\Delta m^2$, i.e.\ with fast 
(averaged) oscillations.%
\footnote{The recent data from the CCFR/NuTeV \protect\cite{CCFR}
experiment and
the preliminary results from the NOMAD experiment \protect\cite{NOMA}
put, however,
upper bounds on $\Delta m^2$ in the same LSND  channel
($\sim 25$ and $\sim 10$ eV$^2$,
respectively).}
The evidence ${\rm S}'$ constrains $\Delta m^2/{\rm eV}^2$ to be either of
${\cal O}(10^{-5})$ \cite{Fo96} or ${\cal O}(10^{-10})$ \cite{Kr95}, 
according to the  specific solution called to explain the $E$-dependence of the
solar $\nu$ deficit [the  Mikheyev-Smirnov-Wolfenstein (MSW) mechanism 
\cite{Wo78}  or ``just-so'' vacuum oscillations \cite{Gl87}, 
respectively]. The evidence ${\rm A}'$ requires 
$\Delta m^2\sim{\cal O}(10^{-2})$ eV$^2$ \cite{Fu94,Fo97}.
The weak evidence ${\rm L}'$ essentially excludes that the LSND detector 
position coincides with the  ``oscillation nodes,'' roughly
corresponding to multiple integers of 4.3 eV$^2$ in $\Delta m^2$
\cite{LSND,Li97}. It should be noted that the evidences in Table~II 
implicitly require the validity of the corresponding evidences in Table~I, 
or, equivalently,  that
\begin{equation}
{\rm X} {\rm\ excluded}\Rightarrow {\rm X}' {\rm\ excluded}\ \ \ \ 
({\rm X}={\rm S},\,{\rm A},\,{\rm or\ L})\ .
\end{equation}

	In the case of three-flavor oscillations,  two independent mass 
square differences, $\delta m^2$ and $m^2$, are involved. 
The $\Delta m^2$ constraints in Tables~I and II apply at least to one of them.

\subsection{Three-family neutrino oscillation parameters}

	We adopt the standard  parametrization \cite{PDBR,Ku89} for the
the unitary mixing matrix $U_{\alpha i}$ connecting the  flavor eigenstates 
$\nu_\alpha=\nu_{e,\mu,\tau}$ to the mass eigenstates
$\nu_i=\nu_{1,2,3}$ ($\nu_\alpha=U_{\alpha i}\nu_i$):
\begin{equation}
U_{\alpha i} = \left(\begin{array}{ccccc}
	c_\omega c_\phi 				&& 
	s_\omega c_\phi 				&&  
	s_\phi						\\
	-s_\omega c_\psi - c_\omega s_\psi s_\phi 	&& 
	c_\omega c_\psi - s_\omega s_\psi s_\phi 	&& 
	s_\psi c_\phi					\\
	s_\omega s_\psi - c_\omega c_\psi s_\phi 	&& 
	-c_\omega s_\psi - s_\omega c_\psi s_\phi 	&& 
	c_\psi c_\phi
\end{array}\right)\ ,
\end{equation}
where $c=\cos$, $s=\sin$, and the mixing angles $\omega$, $\phi$, and $\psi$ 
range between 0 and $\pi/2$. Possible CP violation effects are neglected.

	Without loss of generality, we conventionally denote by $\nu_1$, 
$\nu_3$,  the two mass eigenstates separated by the largest mass gap. In 
particular, $\nu_1$ is chosen to be the closest (in mass) to the
intermediate  state $\nu_2$.  In terms of neutrino masses $m_i$, this choice 
is realized  either with $m_1\leq m_2 \leq m_3$ or with  $m_3\leq m_2 
\leq m_1$. We then define  two independent mass square differences:
\begin{equation}
	\delta m^2=|m^2_2-m^2_1|\ ,\;\;m^2=|m^2_3-m^2_2|\ ,
\end{equation}
that, according to the above convention, obey the inequality 
$\delta m^2\leq m^2$. A scenario with comparable square mass differences
 ($\delta m^2\sim m^2$) will be termed {\em non-hierarchical}.
If $\delta m^2$ is significantly smaller than $m^2$, the scenario
will be termed {\em hierarchical}.

\section{Features of 
Hierarchical $(\delta \lowercase{m}^2 <  \lowercase{m}^2)$
and non-hierarchical $(\delta \lowercase{m}^2 \sim  \lowercase{m}^2)$
scenarios}

	In this section we classify all relevant three-flavor scenarios
according to their spectrum of square mass differences $\delta m^2$
and $m^2$, and comment briefly on their phenomenological implications.
It will become evident that non-hierarchical spectra
$(\delta m^2 \sim m^2)$ are {\em a priori\/} disfavored with
respect to hierarchical spectra $(\delta m^2 < m^2)$.

	For any neutrino experiment, there is a range  of neutrino mass square
difference where genuine {\em oscillation\/} effects are most relevant. For 
current solar, atmospheric, and laboratory (accelerator and reactor)
experiments, these ranges happen to be  roughly decoupled as 
$[0,\,10^{-3.5}]$, $[10^{-3.5},\,10^{-1.5}]$,
and $[10^{-1.5},\,\infty]$, respectively (in units of eV$^2$)
\cite{Fo96,Fa97,Li95,Fo97,Fo95,Li97}.
 Therefore, it is useful to introduce the following shorthand notation 
for these indicative $\Delta m^2$ ranges: 
\begin{eqnarray}
\Delta m^2 \sim \Delta m^2_{\rm sun} 
	&\;\;\;\Longleftrightarrow\;\;\;&
		\Delta m^2 \lesssim 10^{-3.5}\;{\rm\ eV}^2\ ,\\
\Delta m^2 \sim \Delta m^2_{\rm atm} 
	&\;\;\;\Longleftrightarrow\;\;\;&
		10^{-3.5}\lesssim \Delta m^2\lesssim 10^{-1.5}\;{\rm\ eV}^2\ ,\\
\Delta m^2 \sim \Delta m^2_{\rm lab} 
	&\;\;\;\Longleftrightarrow\;\;\;&
		\Delta m^2 \gtrsim 10^{-1.5}\;{\rm\ eV}^2\ ,
\end{eqnarray}
where $\Delta m^2$ denotes either $\delta m^2$ or $m^2$.

	Table~III shows a classification of the $(\delta m^2,\,m^2)$ spectra 
in terms of the above ranges. Six scenarios can be identified.
The first three are hierarchical, the last three are non-hierarchical. None 
of the six cases can fit all the evidences for neutrino mixing (S, A, L in 
Table~I) and for neutrino oscillations (${\rm S}'$, ${\rm A}'$, ${\rm L}'$ in
Table~II), as emphasized in  the last column of Table~III that we now discuss.

	In the first three (hierarchical) cases, $\delta m^2$ and $m^2$ are 
chosen in two different ranges, so as to solve  two out of the three ``neutrino
problems''  posed by the solar deficit, the atmospheric anomaly, and the LSND 
event excess.   In the first scenario, one chooses to fit the solar and the 
laboratory neutrino data. However, this case is not compatible with
${\rm A}'$, since $\delta m^2$ and $m^2$ are, respectively, too small or too 
large to produce a zenith-angle dependence of the atmospheric anomaly. The 
second scenario can provide a good fit to solar and atmospheric data, but 
is not compatible with the evidence L, since $\delta m^2$
and $m^2$ are below the LSND sensitivity. The third scenario  can fit well both
atmospheric and laboratory data. However, due to the assumption of large 
values for $\delta m^2$ and $m^2$, the solar neutrino deficit is predicted to 
be energy-independent and the evidence ${\rm S}'$ must be abandoned.

	In the last three, non-hierarchical scenarios,  both $\delta m^2$
and $m^2$ are assumed to be in the same range, thus ``oversolving'' one of the 
three neutrino problems, at the expenses of the other two. In particular,
the 4th and 5th scenarios in Table~III are not compatible with L, for the
same reason as for the 2nd scenario. In addition, in case 4 the evidence A
must be abandoned,  $\delta m^2$ and $m^2$ being too small to produce 
observable effects for atmospheric neutrinos. In case 5, ${\rm S}'$ must be 
abandoned for the same reason as  for case
3. The reader is referred to Refs.~\cite{Ku89} and \cite{Ya96} 
for studies of subcases of the 4th and 5th scenario, respectively.
The 6th (and last) scenario of Table~III is incompatible with ${\rm S}'$ and 
${\rm A}'$,  since  both $\delta m^2$ and $ m^2$ are too large
to produce $E$-dependent ($L$-dependent) effects for solar (atmospheric)
neutrinos. To our knowledge, there is no recent, post-LSND paper addressing 
case 6.

	Since our goal is to identify the three-flavor scenario
that requires the ``minimum sacrifice'' of data,  we will not further consider
the non-hierarchical scenarios 4, 5, and 6 listed in Table~III.
The hierarchical scenarios 1, 2, and 3 will be instead investigated
in detail in the next section.

\section{Grading hierarchical cases} 

	The  analyses of the hierarchical  cases (1, 2, 3 in Table~III) are 
greatly simplified if the calculations are performed at zeroth order in the 
ratio $\delta m^2/m^2$ \cite{Fo94}.  This is usually a good approximation, 
provided that the relevant values of $\delta m^2$ and $m^2$ are separated
by at least an order of magnitude (see \cite{Fo94} and references therein).  
CP violation effects vanish as $\delta m^2/m^2\to 0$, and thus can also be  
neglected. Within this approximation (sometimes called ``one mass scale 
dominance''), we set out to  analyze in detail the three hierarchical
scenarios, after a brief discussion of the neutrino oscillation probabilities.

\subsection{Oscillation probabilities}

	Table~IV reports the functional form of the three-flavor neutrino 
oscillation probabilities $P^{\alpha\beta}=P(\nu_\alpha\to\nu_\beta)$ for 
solar,  atmospheric, and laboratory  neutrinos---at zeroth order in 
$\delta m^2/m^2$---for each of the three hierarchical scenarios 1, 2, and 3. 
In each specific case,  the probability depends only on a subset of the 
full $3\nu$  parameter space $(\delta m^2,\,m^2,\,U_{\alpha i})$.

	For solar neutrinos, $P^{ee}_{\rm sun}$ can take two functional forms,
$P^{ee}_{\rm MSW}$ or $P^{ee}_{\rm vac}$, according to the  specific solution 
adopted for the solar neutrino problem (MSW or vacuum oscillations, 
respectively). The explicit $3\nu$ expressions of $P^{ee}_{\rm MSW}$ and 
$P^{ee}_{\rm vac}$ can be found, for instance, in \cite{Fo96} and \cite{Fa97}, 
respectively. In both cases (MSW and vacuum) and for both the first and the 
second scenario, $P^{ee}_{\rm sun}$ depends only on the parameters
$(\delta m^2,\,U_{e1},\,U_{e2},\,U_{e3})$, which are further constrained by 
unitarity. In the limit of small $U_{e3}$, the expression of 
$P^{ee}_{\rm sun}$ takes a  more familiar two-family form.   
In the limit of large $\delta m^2$ (scenario 3) both $P^{ee}_{\rm MSW}$ and 
$P^{ee}_{\rm vac}$ assume the simple form given in the seventh row of
Table~IV. Useful graphical representations
of the solar $\nu$ parameter space spanned by $(U_{e1},\,U_{e2},\,U_{e3})$
[or, equivalently, by $(\omega,\,\phi)$] at fixed $\delta m^2$
were introduced in \cite{Fo96}. In particular, we will make use 
of the $(\tan^2\omega,\,\tan^2\phi)$ bilogarithmic chart \cite{Fo96}.

	For atmospheric neutrinos, the flavor oscillation probabilities
take a very simple form, as far as MSW effects in the Earth are ignored. 
The inclusion of matter effects  is not decisive for a qualitative 
understanding and, in any case, it does not require
additional mass-mixing parameters in the expression of $P$, besides those 
listed in Table~IV \cite{Fo97}. In the 2nd scenario, the atmospheric $\nu$
parameter space is $(m^2,\,U_{e3},\,U_{\mu3},\,U_{\tau3})$.
A useful graphical representation at fixed $m^2$
is represented by the  $(\tan^2\psi,\,\tan^2\phi)$ bilogarithmic chart 
\cite{Fo97,Fo95}, which can also be used for scenario 1. Finally,
in the third (hierarchical) scenario, $P_{\rm atm}$ depends on the whole 
mixing matrix (besides $\delta m^2$) and no simple graphical representation 
of the parameter space can be given.

	For laboratory (accelerator and reactor) neutrinos, 
the oscillation probabilities in Table~IV are even simpler than for 
atmospheric $\nu$'s, due to the absence of matter effects. The second scenario 
excludes any  oscillation effect in laboratory neutrino beams, since both
$\delta m^2$ and $m^2$ are small by assumption. Notice that
the laboratory $\nu$ parameter space $(m^2,\,U_{e3},\,U_{\mu3},\,U_{\tau3})$ 
can be charted by the variables $(\tan^2\psi,\,\tan^2\phi)$ at fixed $m^2$, 
as  shown in \cite{Fo95,Li97}.

\subsection{Scenario 1: $(\delta m^2,\,m^2)\sim
(\Delta m^2_{\rm sun },\,\Delta m^2_{\rm lab})$}

	In this scenario, the parameter space of laboratory
(accelerator and reactor) neutrino oscillations is spanned by $m^2$ and 
$U^2_{\alpha3}$ (see Table~IV) or, equivalently, by $m^2$, $\phi$, and $\psi$ 
[see Eq.~(2)]. For this case, the constraints provided by all laboratory data, 
including LSND, have been detailedly worked out in \cite{Li97}. In particular, 
we refer to Fig.~4 of \cite{Li97}, which shows the global bounds in the 
$(\tan^2\psi,\,\tan^2\phi)$ plane for representative values
of $m^2$. These bounds identify  two possible solutions at 90\% C.L.,
one at ``large $\phi$'' and the other at ``small $\phi$.''
The ``large $\phi$'' solution corresponds to $U_{e3}\sim 1$, that is,
to $\nu_e$ almost coincident with the mass eigenstate $\nu_3$.
In this case, the deficit of solar neutrinos would be of order
$(1-U^2_{e3})^2$ and thus negligible \cite{Fo96},
in contrast with the evidence S. The ``small $\phi$'' solution is also
characterized by large values of $\psi$ and thus it
corresponds to a $U_{\mu3}\sim 1$, i.e., to  $\nu_\mu$ 
 almost coincident with the mass eigenstate $\nu_3$. In this case,
the remaining flavor states $(\nu_e,\,\nu_\tau)$ have to be mixed prevalently
with the mass eigenstates $(\nu_1,\,\nu_2)$, implying that solar
neutrino oscillations are almost pure  $\nu_e\leftrightarrow\nu_\tau$.

	The above considerations imply that in the scenario 1  all the 
laboratory results (including  the LSND evidences L and ${\rm L}'$) and
all the solar neutrino data (evidences S and ${\rm S}'$) can be reconciled 
by combining the ``small $\phi$'' solution of laboratory $\nu$ fits
\cite{Li97} with almost pure $\nu_e\to\nu_\tau$  oscillations
for solar neutrinos. The $\nu_e\to\nu_\tau$ oscillations can be assumed 
either of the MSW type \cite{Wo78} or of the vacuum 
oscillation type \cite{Gl87}. In the MSW case, the bounds in the 
solar neutrino parameter space $(\delta m^2,\,\omega,\,\phi)$
have been explicitly worked out in \cite{Fo96}. In particular, Fig.~12
in \cite{Fo96} shows that in the limit of large $\phi$ no solution is allowed,
while in the limit of small $\phi$ one recovers the familiar
``small $\omega$'' and ``large $\omega$'' solutions  usually
found in two-family MSW fits.

	It remains to be seen whether the atmospheric data can also be fitted
in this scenario. We have already observed in Table~III that  one cannot fit 
the evidence ${\rm A}'$ in this case. Indeed, the global three-flavor
fits to the atmospheric $\nu$ data (including the evidences A and ${\rm A}'$)
\cite{Fo97} and to the laboratory $\nu$ data \cite{Li97} favor two different, 
incompatible ranges for $m^2$. However, it has been noted by Cardall and 
Fuller \cite{Ca97} that the sole exclusion of the evidence ${\rm A}'$ allows 
the laboratory and the  atmospheric data to be (marginally)
reconciled at $m^2\sim 0.4$ eV$^2$. It must be said that
the evidence ${\rm A}'$ is somewhat uncertain.  Its statistical significance 
might be smaller \cite{Li95,Sa95} than the value
originally quoted by the Kamiokande collaboration \cite{Fu94}. 
Moreover,  ${\rm A}'$ is not confirmed by a recent reanalysis of older
data from the IMB experiment \cite{Cl97}.  The SuperKamiokande experiment 
\cite{Yo97} is expected to clarify soon this issue.

	Motivated by the semiquantitative results of \cite{Ca97} we have 
repeated a global fit to all atmospheric data as in \cite{Fo97}, but with 
{\em unbinned\/} multi-GeV Kamiokande data, so as to integrate out their 
zenith-angle dependence (evidence ${\rm A}'$). The results of this new fit 
are reported in Fig.~1. Figure~1 shows, in the $(\tan^2\psi,\,\tan^2\phi)$ 
plane and for the representative value  $m^2=0.45$ eV$^2$,
the regions allowed at 90\% C.L.\ by fits to all laboratory data
(dashed line) and to all sub-GeV and unbinned multi-GeV atmospheric data
(dotted line). It can be seen that these two regions overlap, although
marginally, at $(\tan^2\psi,\,\tan^2\phi)\sim(7,\,0.0035)$.  Therefore,
it makes sense to perform a combined fit. The resulting solution
is shown as a solid contour in Fig.~1. Its area is larger than the
simple intersection of the previous two regions, as is the
case for the combination of data that are marginally compatible;
therefore, its  shape  must be taken with caution. 
Actually, this solution appears only in the narrow range 
$0.3\lesssim m^2\lesssim 0.5$ eV$^2$, as also noticed in \cite{Ca97}.

	Figure~2 represents a concise summary of the global fit(s)
to neutrino oscillation data (excluding ${\rm A}'$) in the three-flavor
scenario 1. The $(\tan^2\psi,\,\tan^2\phi)$ plane on the left (which reports 
a schematic reduction of Fig.~1) is basically the parameter space of the 
neutrino state $\nu_3$ \cite{Fo95,Li97}.  As already noticed,
the ``terrestrial'' (laboratory and atmospheric) data impose
$(\tan^2\psi,\,\tan^2\phi)\sim(7,\,0.0035)$. This  solution survives with 
similar characteristics in the range $0.3\lesssim m^2\lesssim 0.5$ eV$^2$. The
$(\tan^2\omega,\,\tan^2\phi)$ plane on the right of Fig.~2
maps the parameter space of solar $\nu_e$'s. For the small
value of $\phi$  fixed by the terrestrial $\nu$ fit, solar neutrino
oscillations are almost pure $\nu_e\leftrightarrow\nu_\tau$. 
Three solutions are possible: (1a) small angle MSW; (1b) large angle MSW; and 
(1c) vacuum oscillations. The mass and mixing parameters corresponding to
these three possible solutions are reported in Table~V. The given $\tan^2$ 
values  may vary within a factor of about two, as reminded by the
 ``thickness'' of the black dots in Fig.~2. Finally, the flavor components of 
$\nu_3$ and the mass components of $\nu_e$ are explicitly worked out
at the bottom of Fig.~2, for the mixing values reported in Table~V.

	In conclusion, terrestrial and solar neutrino data can all be fitted
within the three-flavor scenario 1,  with the only exclusion of the 
zenith-angle dependence of the atmospheric $\nu$ anomaly (evidence ${\rm A}')$.
The resulting constraints in the space  
$(\delta m^2,\,m^2,\,\omega,\,\phi,\,\psi)$ are rather stringent
for each of the three possible solutions corresponding to different
solar $\nu$ fits. As we will see, the two remaining hierarchical scenarios 2
and 3 demand more drastic exclusions of data. In this sense, 
the scenario 1 requires the ``minimum sacrifice.'' Its implications
will be studied  in Sec.~V.

\subsection{Scenario 2: $(\delta m^2,\,m^2)\sim
(\Delta m^2_{\rm sun },\,\Delta m^2_{\rm atm})$}

	This case requires the sacrifice of the whole LSND data set 
(evidence L) as noted in Table~III and related comments. 
In view of the fact that new, independent LSND data from pion decay in flight 
\cite{Lnew}  have recently  confirmed the existing LSND data from muon decay 
at rest \cite{LSND}, this appears a more drastic rejection than the 
exclusion of ${\rm A}'$ in the previously analyzed scenario 1.

	In scenario 2, one has to combine the {\em negative\/} laboratory 
oscillation searches with the atmospheric and solar neutrino data.
The complete three-flavor analysis of atmospheric neutrino data
performed in \cite{Fo97} applies to this case. The bounds on 
the $(m^2,\,\psi,\,\phi)$ parameter space are given in Fig.~7
of Ref.~\cite{Fo97} for atmospheric data only, and in Fig.~11 of the same 
paper for atmospheric and laboratory data (without LSND). 
Concerning the MSW solution to the solar neutrino problem, the 
$(\delta m^2,\,\omega,\,\phi)$ bounds
appropriate to this scenario can be read from Fig.~12 in \cite{Fo96}.%
\footnote
{An analogous, detailed $3\nu$ analysis is still lacking for the vacuum 
oscillation solution to the solar neutrino problem.
See, however, \protect\cite{Be96} for a recent $3\nu$  fit.}
The match between solar and terrestrial bounds requires  that the common 
parameter $\phi$ assume the same value in both fits. This exercise is made 
easier by the fact that the panels of Fig.~12 in \cite{Fo96} and the
panels of Figs.~7 and 11 in \cite{Fo97} exhibit purposely
the same values of $\tan^2\phi$ on the $y$-axis.

	The so-called ``threefold maximal mixing'' framework
of Harrison, Perkins, and Scott \cite{Ha97} is a subcase of this scenario. 
It corresponds to $|U_{\alpha i}|^2=1/3$, i.e., to 
$(\tan^2\omega,\,\tan^2\phi,\,\tan^2\psi)=(1,\,1/2,\,1)$ (at any value of 
$\delta m^2$, $m^2$). The point $(\tan^2\phi,\,\tan^2\psi)=(1/2,\,1)$ is 
allowed by terrestrial $\nu$  data, provided that $m^2$ is below the 
sensitivity of the established laboratory experiments (that do not allow large
neutrino mixing) \cite{Fo95}. However,  the point 
$(\tan^2\omega,\,\tan^2\phi)=(1,\,1/2)$ is excluded by the MSW analysis of 
solar $\nu$ data at more than 99\% C.L.\ \cite{Fo96}. The same point might be 
instead allowed by the vacuum oscillation solution of solar $\nu$ data, as 
suggested in \cite{Ha97},  although a detailed  three-flavor analysis of 
this situation has not been performed yet.

\subsection{Scenario 3: $(\delta m^2,\,m^2)\sim
(\Delta m^2_{\rm atm },\,\Delta m^2_{\rm lab})$}

	This scenario has been recently studied in \cite{Mi95,Go95} and, in 
particular, by Acker and Pakvasa in \cite{Ac96}. It is particularly appealing 
to  researchers  of laboratory neutrino  oscillations, since 
$\Delta m^2_{\rm lab}$ represents the range explorable by present and future 
short baseline experiments, and $\Delta m^2_{\rm atm}$ covers the range 
explorable by future long baseline experiments.  At first sight, this 
scenario seems to require only the exclusion of the  evidence ${\rm S}'$ 
(see Table~III). However, a more detailed investigation shows that 
additional sacrifices of data are necessary.

	Let us consider first the chain of constraints imposed by the
{\em negative\/} oscillation searches at accelerators and reactors
in the parameter space $(m^2,\,U_{\alpha3})\equiv (m^2,\,\phi,\,\psi)$,
as spanned by $\nu_3$ \cite{Fo95}. The LSND data will be reintroduced at 
the end.   The negative laboratory   searches in all channels
constrain $\nu_3$ to be  close to one of the mass eigenstates,
in order to suppress the (unobserved) oscillation phenomena.
In Table~VI, the three possible cases $\langle\nu_3|\nu_\alpha\rangle\sim 1$
with $\alpha=e$, $\mu$, or $\tau$, are labeled 3a, 3b, and 3c, respectively.
For each case, the resulting (approximate) structure of the 
mixing matrix  is given in the third column, where
the parameters $a$, $b$, and $c$ are {\em a priori\/} unconstrained. 
The fourth column in Table~VI gives the expression of
the solar neutrino survival probability, as derived from Table~IV for
the scenario 3 and its subcases 3a, 3b, and 3c. Notice
that for $\langle\nu_3|\nu_e\rangle\sim 1$ (case 3a) the states
$\nu_1$ and $\nu_2$ must be essentially a mixture of $\nu_\mu$ and
$\nu_\tau$. Therefore, in case 3a  the atmospheric neutrino oscillations are
almost pure $\nu_\mu\leftrightarrow\nu_\tau$, with an effective
mixing amplitude $\sin^2\theta^{\rm eff}_{\mu \tau}$ equal to
$4a(1-a)$, as given in the last column of Table~VI. Similar
considerations apply to cases 3b and 3c. In other words,
atmospheric neutrino oscillations are governed in each case
by the only nontrivial $2\times 2$ submatrix of the mixing matrix.
We remark that the properties listed in the third, fourth and fifth
column of Table VI follow solely from the assumption in the
first column of the same Table, i.e.\ from the negative results of
laboratory oscillation searches.

	From Table~VI it follows that the case 3a is compatible
with the atmospheric evidences A and ${\rm A}'$,  fitted with almost pure 
$\nu_\mu\leftrightarrow\nu_\tau$ oscillations. However, this subcase predicts 
$P^{ee}_{\rm sun}\sim 1$ and thus no neutrino deficit. Conversely, in case 3b 
one can obtain  a solar neutrino deficit by tuning the free parameter $b$, but 
then aggravates (instead of solving) the atmospheric neutrino anomaly 
through  $\nu_e\leftrightarrow\nu_\tau$ oscillations.

	Only the case 3c seems to survive, since with a judicious choice of 
the free  parameter $c$ (Table~III) one can obtain both a significant deficit 
of solar neutrinos and a solution to the atmospheric anomaly through
$\nu_e\leftrightarrow\nu_\mu$ oscillations. (It is  precisely this scenario 
that was studied in \cite{Ac96}.)  As far as the LSND data are excluded, it  
seems to provide a reasonable fit to all data (excluding, of course, the 
evidence ${\rm S}'$).  However, as shown in \cite{Li97}, the inclusion of the 
LSND results in the fit to laboratory $\nu$ data strongly disfavors the 
situation $\langle \nu_3|\nu_\tau\rangle\sim 1$, i.e.\ the case 3c. In fact, 
the two possible solutions found at 90\% C.L.\ in \cite{Li97} at any given 
$m^2$ (see also Fig.~1 of this work) correspond to cases 3a and 3b, 
respectively. The semiquantitative analyses in \cite{Mi95,Go95,Ac96} and 
\cite{Ba95}  found an allowed solution corresponding to  case 3c only by 
``stretching'' all the uncertainties of the laboratory data (including LSND) 
to their 90\% C.L. limits.%
\footnote{In our notation, Ref.~\protect\cite{Ac96,Ba95} studied
the case 3c, Ref.~\protect\cite{Mi95} the cases 3a and 3c, and
Ref.~\protect\cite{Go95} the cases 3a, 3b, and 3c.}
However, the overall C.L.\ of this contrived situation  certainly exceeds
 90\%. Indeed, a proper statistical analysis  excludes the case 3c at 
$\sim 99\%$ C.L.\ \cite{Li97}. A similar conclusion about the case 3c
was independently reached in \cite{Bi96}.

	In conclusion, the scenario 3 is incompatible not only with ${\rm S}'$,
but also with other evidences for neutrino oscillations, depending
on the specific subcase studied (3a, 3b, or 3c in Table~VI).
The claim that in this framework ``three neutrino flavors are enough'' to
fit the world neutrino data \cite{Ac96} is not substantiated by a 
quantitative analysis---one has to discard at least a primary 
evidence for mixing (S, A, or L) in order to achieve a reasonable fit.
A summary of the data not compatible with scenario 3 is given in Table~VII.

\section{Implications of the ``minimum sacrifice'' scenario} 

	In this section we discuss the implications of the scenario 1 that,
at present, emerges as the three-flavor framework consistent with
the largest amount of neutrino data. Only the zenith-angle dependence of
the multi-GeV Kamiokande data is incompatible with this scenario.
Therefore, an immediate ``falsification'' of this framework could
be provided by the high-statistics SuperKamiokande experiment, should
it confirm the zenith-angle dependence of the 
atmospheric $\nu$ anomaly found in kamiokande \cite{Fu94}.
Preliminary results from SuperKamiokande \cite{Yo97} confirm the anomaly
of the {\em total\/} rates, but do not provide yet any decisive
indication on the zenith-angle spectrum.

	Assuming that the scenario 1 survives the SuperKamiokande test,
which are its implications on solar and laboratory neutrino oscillation
searches? As discussed in Sec.~IV~B, solar neutrino oscillations are predicted 
to be almost pure $\nu_e\leftrightarrow\nu_\tau$ in this case. 
Unfortunately, since solar neutrino experiments do not distinguish 
$\nu_\mu$'s from $\nu_\tau$'s  in solar neutrino experiments,  we do not 
expect important indications on this scenario from the new-generation 
solar $\nu$ detectors SuperKamiokande \cite{Yo97} or Sudbury
Neutrino Observatory (SNO) \cite{Hi96}.

	Short-baseline laboratory oscillations should, instead, give
important information. Figure~3 shows the solution corresponding to scenario 
1 (thick, solid line), together with the regions of the mixing parameter space
that will be probed by the KARMEN experiment \cite{KARM} (exploring the same 
LSND channel $\bar\nu_\mu\leftrightarrow\bar\nu_e$) at the end of 1999, by 
the CERN experiments CHORUS \cite{CHOR} and NOMAD \cite{NOMD} 
($\nu_\mu\leftrightarrow\nu_\tau)$ in four years of data taking, and by the 
proposed $\nu_\mu\leftrightarrow\nu_\tau$ experiments  COSMOS \cite{COSM}
at Fermilab and TENOR/NAUSICAA \cite{TENA} or TOSCA \cite{TOSC} at CERN.
It can be seen that these experiments can probe either a fraction or all
of the proposed solution. It is particularly interesting to note that,
although this solution is governed by the results of a
$\bar\nu_\mu\leftrightarrow\bar\nu_e$ experiment (LSND),
it can be completely covered by sensitive
$\nu_\mu\leftrightarrow\nu_\tau$ experiments (COSMOS, TOSCA, or 
TENOR/NAUSICAA). This feature would have  gone unnoticed in a two-flavor 
analysis.

	Concerning long-baseline experiments, the current reactor projects
Chooz \cite{Choo} and Palo Verde \cite{Palo}  have not sufficient sensitivity 
in mixing to probe the solution in Fig.~3, and, in general, the
LSND signal \cite{Li97}. In accelerator experiments,
the solution of Fig.~3 would induce oscillations with  a wavelength much 
smaller than the  (long) baseline. The finite
energy bandwidth of the beam and of the detector would then
smear out the oscillation pattern, giving rise to evidences
of {\em mixing\/} at best.  This would, in a sense, defeat main purpose and 
hope of long-baseline experiments.

\section{Conclusions}

	We have analyzed how well various possible three-flavor scenarios
can fit the world neutrino data from oscillation searches in solar
and terrestrial (atmospheric and laboratory) experiments. In any case
some data must be ``sacrificed.'' Scenarios with no hierarchy 
in the spectrum of neutrino mass differences are shown to be {\em a
priori\/} disfavored.  A specific hierarchical
framework emerges naturally as the ``minimum sacrifice'' fit to the data.
The characteristics and the phenomenological consequences of this scenario 
have been investigated. If this scenario survives the stringent test  of
the SuperKamiokande atmospheric measurements, then future
short-baseline searches at accelerator facilities will play a decisive
role in its (dis)confirmation.

\acknowledgments

	One of us (G.L.F.) thanks the organizers of the 
Workshop on Fixed Target Physics at the Fermilab Main Injector, Fermilab,
Batavia IL, where preliminary results of this work were presented.



\begin{table}
\caption{	Evidences for neutrino {\em mixing}\/ and their implications
		on $\Delta m ^2$.}
\begin{tabular}{lcl}
Primary evidence (for $\nu$ {\em mixing})&Symbol& 
					Implications for $\Delta m^2$\\
\tableline
Solar $\nu_e$ deficit 			& S &
				$\Delta m^2\gtrsim 10^{-11}$ eV$^2$\\
Atmospheric $\nu_\mu/\nu_e$ anomaly 	& A &
				$\Delta m^2\gtrsim 10^{-4}$  eV$^2$\\
LSND $\bar\nu_e$ signal 		& L &
				$\Delta m^2\gtrsim 10^{-1}$  eV$^2$\\
\end{tabular}
\end{table}

\vfill

\begin{table}
\caption{	Evidences for neutrino {\em oscillations}\/ and their 
		implications on $\Delta m ^2$.}
\begin{tabular}{lcl}
Additional evidence (for $\nu$ {\em oscillation})&Symbol&
					Implications for $\Delta m^2$\\
\tableline
$E$-dependence of solar $\nu$ deficit 		&${\rm S}'$&
	\begin{tabular}{l}
	$\Delta m^2\sim10^{-5}$ eV$^2$\ (MSW)\ or\\
	$\Delta m^2\sim 10^{-10}$ eV$^2$\ (vacuum)\end{tabular}\\
$L$-dependence of atmospheric $\nu$ anomaly 	&${\rm A}'$&
	$\Delta m^2\sim 10^{-2}$ eV$^2$\\
$(L/E)$-dependence of LSND $\nu$ signal 	&${\rm L}'$&
	$\Delta m^2\not\simeq n\times4.3$ eV$^2$ $(n=1,\,2,\,3,\dots)$
\end{tabular}
\end{table}

\vfill

\begin{table}
\caption{Classification of $3\nu$ scenarios according to their spectrum of 
	square mass differences $(\delta m^2,\,m^2)$, as compared with the 
	phenomenologically relevant $\Delta m^2$ ranges. The first three 
	scenarios are ``hierarchical'' $(\delta m^2 < m^2)$, the last three
	``non-hierarchical'' $(\delta m^2 \sim m^2)$. The last column shows 
	the (minimum) set of evidences incompatible with the various scenarios.}
\begin{tabular}{ccccc}
&\multicolumn{3}{c}{Ranges of $\nu$  square mass difference (eV$^2$)} & \\ 
\cline{2-4} & 
$\Delta m^2_{\rm sun}$ & 
$\Delta m^2_{\rm atm}$ & 
$\Delta m^2_{\rm lab}$ & 
Evidences to be   							\\
Scenario  & 
$\lesssim10^{-3.5}$ & 
$\sim10^{-3.5}$--$10^{-1.5}$ & 
$\gtrsim 10^{-1.5}$ & 
discarded (at least)							\\
\tableline
1&$\delta m^2$ 	     & 			 &$m^2$ 	      		&
						${\rm A}'$		\\
2&$\delta m^2$ 	     &$m^2$ 		 & 		      		&
						${\rm L}$		\\
3&		     &$\delta m^2$ 	 &$m^2$ 	      		&
						${\rm S}'$  		\\
4&$\delta m^2$, $m^2$& 			 & 		      		&
						${\rm L}$ and ${\rm A}$	\\
5&		     &$\delta m^2$, $m^2$& 		      		&
						${\rm L}$ and ${\rm S}'$\\
6&		     & 			 &$\delta m^2$, $m^2$		&
						${\rm A}'$ and ${\rm S}'$\\
\end{tabular}
\end{table}

\newpage

\begin{table}
\squeezetable
\caption{	Functional form of the oscillation probabilities
		for solar, atmospheric, and laboratory neutrino beams, in
		each of the hierarchical scenarios 1, 2, and 3.}
\begin{tabular}{cccl}
Scenario & $P$ & Parameters &  Functional form of the oscillation
probability\tablenotemark[1] \\
\tableline
& 	$P_{\rm sun}^{ee}$&$(\delta m^2,\,U_{ei})$&$
  	P_{\rm MSW}^{ee}(\delta m^2,\,U_{ei})$\ \ \ or \ \ $
  	P_{\rm vac}^{ee}(\delta m^2,\,U_{ei})$  			\\
1 	&$P^{\alpha\beta}_{\rm atm}$&$(U_{\alpha3},\,U_{\beta3})
	$&$\delta_{\alpha\beta}-2U_{\alpha3}U_{\beta3}(\delta_{\alpha\beta}-
	U_{\alpha3}U_{\beta3})$ + {\em matter effects}			\\
&	$P^{\alpha\beta}_{\rm lab}$&$(m^2,\,U_{\alpha3},\,U_{\beta3})$
	&$\delta_{\alpha\beta}-4U_{\alpha3}U_{\beta3}(\delta_{\alpha\beta}-
	U_{\alpha3}U_{\beta3})\sin^2(1.27 m^2 L/E)$			\\
\tableline
& 	$P_{\rm sun}^{ee}$&$(\delta m^2,\,U_{ei})$&$
  	P_{\rm MSW}^{ee}(\delta m^2,\,U_{ei})$\ \ \ or \ \ $
  	P_{\rm vac}^{ee}(\delta m^2,\,U_{ei})$   			\\
2 	&$P^{\alpha\beta}_{\rm atm}$&$
  	(m^2,\,U_{\alpha3},\,U_{\beta3})$&$\delta_{\alpha\beta}-
  	4U_{\alpha3}U_{\beta3}(\delta_{\alpha\beta}-
	U_{\alpha3}U_{\beta3})\sin^2(1.27 m^2 L/E)
  	$ + {\em matter effects}					\\
& 	$P^{\alpha\beta}_{\rm lab}$&---&$\delta_{\alpha\beta}$		\\
\tableline
&	$P_{\rm sun}^{ee}$&$(U_{ei})$&$
  	1-2(U^2_{e1}U^2_{e2}+U^2_{e2}U^2_{e3}+U^2_{e3}U^2_{e1})$ 	\\
3 	&$P^{\alpha\beta}_{\rm atm}$&$
  	(\delta m^2,\,U_{\alpha i},\,U_{\beta i})$&$\delta_{\alpha\beta}-
  	2U_{\alpha3}U_{\beta3}(\delta_{\alpha\beta}-
	U_{\alpha3}U_{\beta3})-4U_{\alpha1}U_{\alpha2}U_{\beta1}U_{\beta2}
	\sin^2(1.27 \delta m^2 L/E)$ + {\em matter effects}		\\
& 	$P^{\alpha\beta}_{\rm lab}$&$(m^2,\,U_{\alpha3},\,U_{\beta3})$
	&$\delta_{\alpha\beta}-4U_{\alpha3}U_{\beta3}(\delta_{\alpha\beta}-
	U_{\alpha3}U_{\beta3})\sin^2(1.27 m^2 L/E) $%
\end{tabular}
\tablenotetext[1]{\scriptsize
Units: $[\delta m^2],\,[m^2]={\rm eV}^2$, $[L]={\rm m}$, $[E]={\rm MeV}$.}
\end{table}

\vfill

\begin{table}
\caption{Scenario 1: Summary of the neutrino mass-mixing parameters 
	for the three possible subcases 1a, 1b, and 1c. Mixings are
	determined up to a factor of about two.}
\begin{tabular}{ccccccc}
Scenario  & $m^2$ (eV$^2$) &
$\tan^2\psi$ & $\tan^2\phi$ & Solar neutrino solution &
$\delta m^2$ (eV$^2$) & $\tan^2\omega$\\
\tableline
1a&$\sim0.3$--$0.5$&$\sim7$&$\sim0.0035$&MSW, small $\omega$&
$\sim 0.4$--$1\times10^{-5}$ &$\sim0.0015$ \\
1b&$\sim0.3$--$0.5$&$\sim7$&$\sim0.0035$&MSW, large $\omega$&
$\sim 0.6$--$9\times10^{-5}$ &$\sim0.25$ \\
1c &$\sim0.3$--$0.5$&$\sim7$
&$\sim0.0035$&vacuum oscillation &$\sim 0.5$--$1\times10^{-10}$
&$\sim 1$ 
\end{tabular}
\end{table}

\vfill

\begin{table}
\caption{Scenario 3: Implications of negative oscillation searches for
	solar and atmospheric neutrino oscillations. See the text for details.}
\begin{tabular}{ccccc}
 						& 
$\nu_3$ from negative 				& 
Approximate  					& 
Implications for 				& 
Implications for 					\\
Scenario 						& 
lab.\ searches 					& 
structure of $U^2_{\alpha i}$  			& 
solar neutrinos					& 
atmospheric neutrinos 					\\ 
\tableline
\noalign{\vskip5pt}
3a						& 
$\langle\nu_3 | \nu_e  \rangle\sim 1$ 		& 
$\left(\begin{array}{ccc}
0   & 0   & 1   \\
1-a & a   & 0   \\
a   & 1-a & 0 
\end{array}\right)$				& 
$P^{ee}_{\rm sun}\sim 1$			& 
\begin{tabular}{c}
$\sim (\nu_\mu,\,\nu_\tau)$ oscillations,\\
$\sin^22\theta_{\mu\tau}^{\rm eff}\sim4a(1-a)$
\end{tabular}						\\
\noalign{\vskip5pt}
3b						& 
$\langle\nu_3 | \nu_\mu \rangle\sim 1$ 		& 
$\left(\begin{array}{ccc}
1-b & b   & 0   \\
0   & 0   & 1   \\
b   & 1-b & 0 
\end{array}\right)$				& 
$P^{ee}_{\rm sun}\sim 1-2b(1-b)$		& 
\begin{tabular}{c}
$\sim (\nu_e,\,\nu_\tau)$ oscillations,\\
$\sin^22\theta_{e\tau}^{\rm eff}\sim4b(1-b)$
\end{tabular}						\\
\noalign{\vskip5pt}
3c					& 
$\langle\nu_3 | \nu_\tau\rangle\sim 1$ 		& 
$\left(\begin{array}{ccc}
1-c & c   & 0   \\
c   & 1-c & 0   \\
0   & 0   & 1 
\end{array}\right)$				& 
$P^{ee}_{\rm sun}\sim 1-2c(1-c)$		& 
\begin{tabular}{c}
$\sim (\nu_e,\,\nu_\mu)$ oscillations,\\
$\sin^22\theta_{e\mu}^{\rm eff}\sim4c(1-c)$
\end{tabular}						\\
\end{tabular}
\end{table}

\newpage

\begin{table}
\caption{Scenario 3: 
Summary of the evidences to be discarded in each of the three subcases 
3a, 3b, and 3c.}
\begin{tabular}{cc}
		& Incompatible \\
Scenario 	& evidence(s)  \\
\tableline
3a		& S		\\
3b		& ${\rm S}'$, A	\\
3c		& ${\rm S}'$, L		
\end{tabular}
\end{table}



\begin{figure}
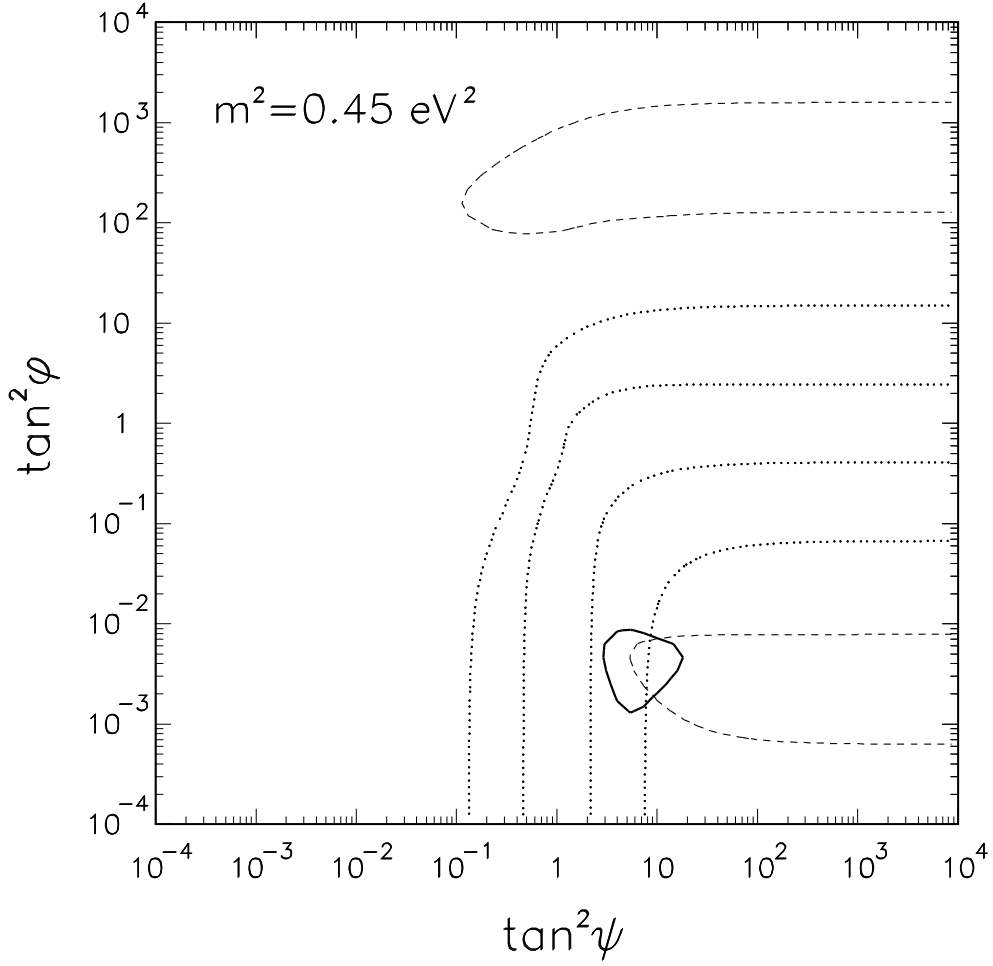

\caption{Results of the fit to terrestrial neutrino data 
	(90\% C.L.) in the scenario 1,
	for $m^2=0.45$ eV$^2$. Results for $0.3\protect\lesssim m^2 
	\protect\lesssim 0.5$ eV$^2$ (not shown) are similar.}
\end{figure}
\begin{figure}
\caption{Schematic results of the combined fit to solar and terrestrial	
	neutrino data in the scenario 1.}
\end{figure}
\begin{figure}
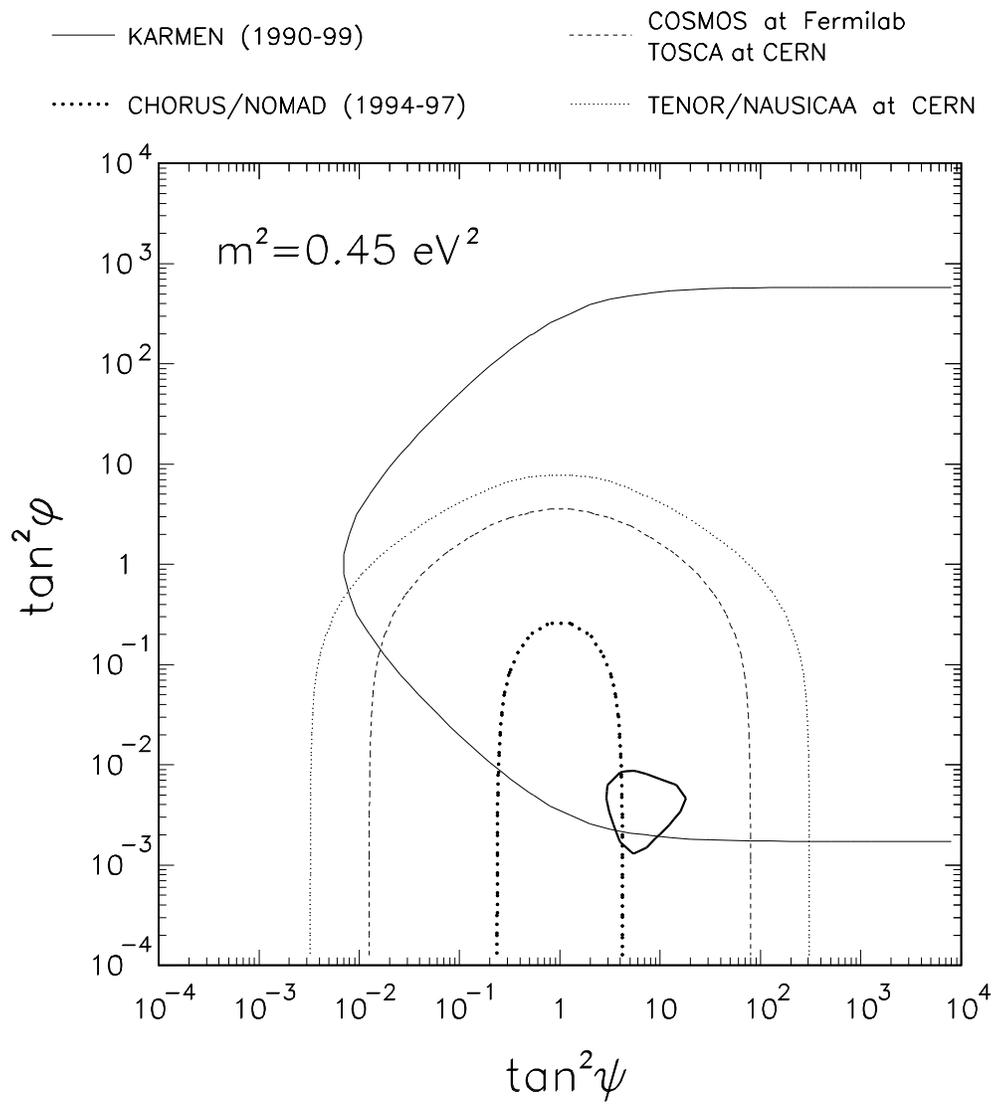

\caption{Implications	of the scenario 1 for running and future
	short-baseline experiments.}
\end{figure}


\newcommand{\InsertFigure}[2]{\newpage\begin{center}\mbox{%
\epsfig{bbllx=1.4truecm,bblly=1.3truecm,bburx=19.5truecm,bbury=26.5truecm,%
height=21.truecm,figure=#1}}\end{center}\vspace*{-1.85truecm}%
\parbox[t]{\hsize}{\small\baselineskip=0.5truecm\hskip0.5truecm #2}}

\InsertFigure{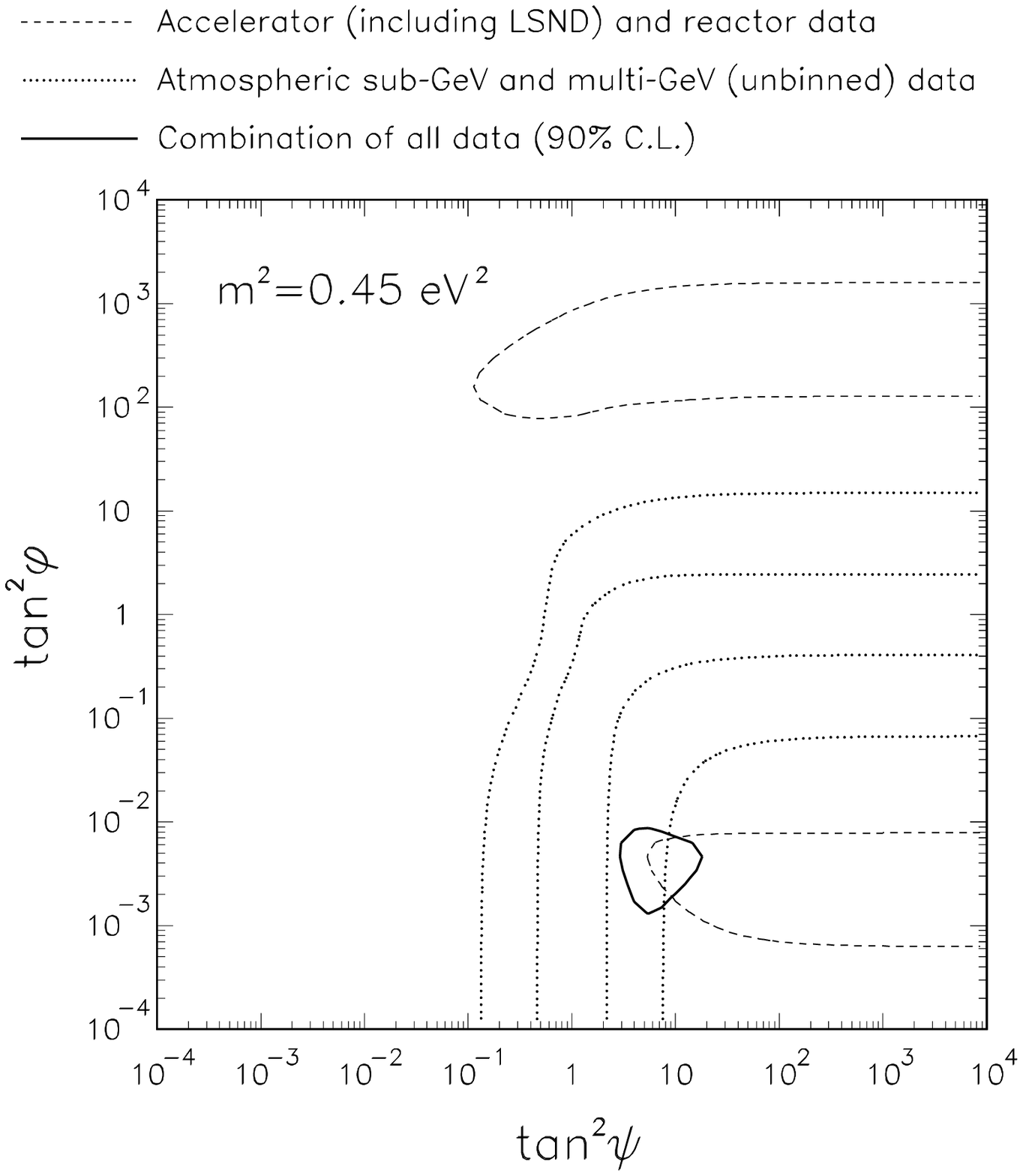}%
{FIG.~1. Results of the fit to terrestrial neutrino data 
	(90\% C.L.) in the scenario 1,
	for $m^2=0.45$ eV$^2$. Results for $0.3\lesssim m^2 \lesssim 0.5$
	eV$^2$ (not shown) are similar.}
\InsertFigure{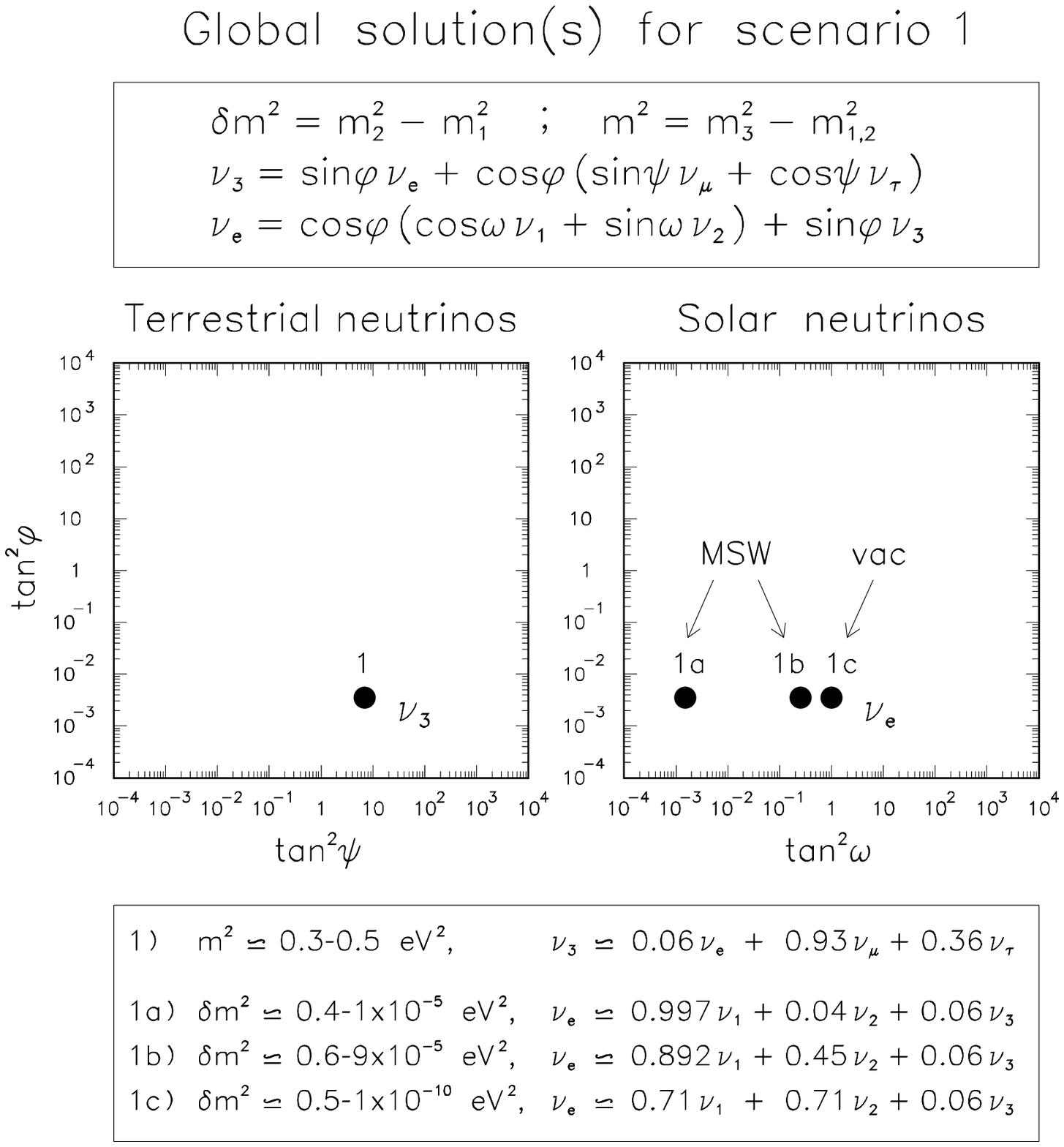}%
{FIG.~2. Schematic results of the combined fit to solar and terrestrial	
	neutrino data in the scenario 1.}
\InsertFigure{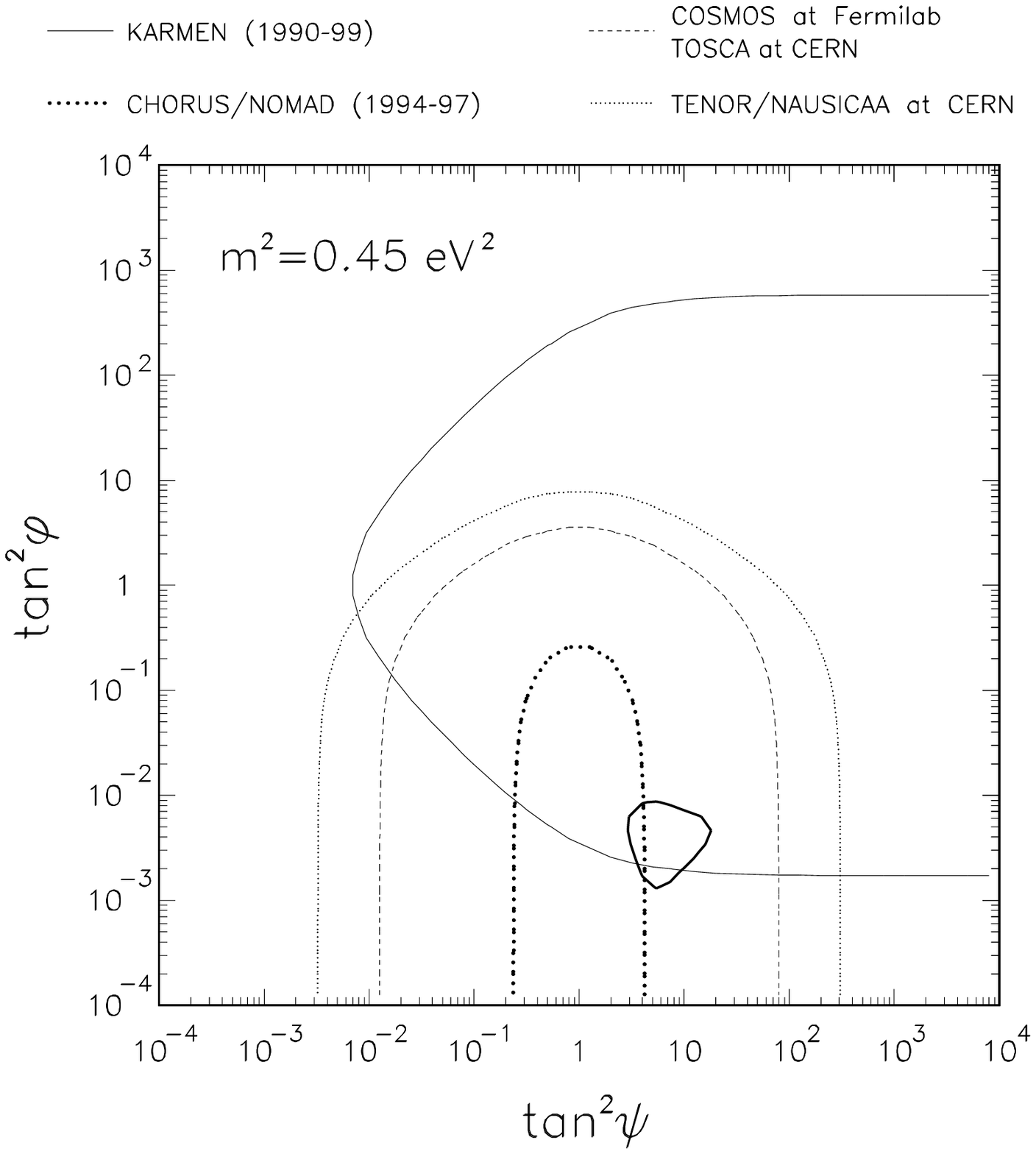}%
{FIG.~3. Implications	of the scenario 1 for running and future
	short-baseline experiments.}

\end{document}